\documentclass[sn-aps]{sn-jnl}% American Physical Society (APS) Reference Style

\usepackage{amsmath}
\usepackage{wrapfig}
\usepackage{bm}
\usepackage{subcaption}
\usepackage{overpic}
\usepackage{placeins}
\usepackage{lipsum}
\usepackage{tikz}
\usetikzlibrary{shapes.geometric}
\usetikzlibrary{intersections}
\usetikzlibrary{calc}
\usetikzlibrary{decorations.pathreplacing}
\usepackage{program}

\jyear{2022}%

\begin{document}

\title[Impact of Engine Nacelle Flow on Buffet]{Impact of Engine Nacelle Flow on Buffet}

\author*[1]{\fnm{Thomas} \sur{L\"urkens}}\email{t.luerkens@aia.rwth-aachen.de}

\author[1,2]{\fnm{Matthias} \sur{Meinke}}\email{m.meinke@aia.rwth-aachen.de}

\author[1,2]{\fnm{Wolfgang} \sur{Schr\"oder}}\email{office@aia.rwth-aachen.de}

\affil[1]{\orgdiv{Chair of Fluid Mechanics and Institute of Aerodynamics}, \orgname{RWTH Aachen University}, \orgaddress{\street{W\"ullnerstra{\ss}e 5a}, \city{Aachen}, \postcode{52062}, \country{Germany}}}

\affil[2]{\orgdiv{JARA Center for Simulation and Data Science}, \orgname{RWTH Aachen University}, \orgaddress{\street{Seffenter Weg 23}, \city{Aachen}, \postcode{52074}}}

%%==================================%%
%% sample for unstructured abstract %%
%%==================================%%

\abstract{
The transonic flow around the OAT15A airfoil is computed at buffet conditions, i.e., freestream Mach number $Ma{_\infty} = 0.73$, chord-based freestream Reynolds number $Re_c = 2\cdot10^6$, and angle of attack $\alpha = 3.5^\circ$ using wall-modeled LES. Two configurations are considered, one which includes a generic ultra-high bypass ratio (UHBR) engine nacelle geometry and one without an engine, which is denoted the baseline case. The introduction of the UHBR-engine nacelle leads to a significant deviation of the flow onto the airfoil from the baseline case and has an essential effect on the occurring shock dynamics. The flow field of the nacelle configuration is characterized by a shock wave on the upper part of the nacelle sharing dynamic features with the shock on the airfoil. This impact of the nacelle shock on the airfoil shock means a reduced strength of the airfoil shock resulting in a less developed buffet. The perturbation of the general flow field is evaluated as to established buffet models and the dynamic features of the shock waves are analyzed by the sparsity-promoting dynamic mode decomposition. This analysis shows the existence of a shared dynamic mode of the nacelle and the airfoil shock which suggests a coupling mechanism between them.}

\keywords{Transonic Buffet, Engine Nacelle Flow, Wall-Modeled LES}

\maketitle

\section{Introduction}
In transonic flow on wings of transport aircraft a local supersonic region is formed that is terminated by a shock wave. The buffet phenomenon is characterized by a self-sustained large-scale oscillation of this shock wave at low frequencies in the range of $Sr \sim \mathcal{O}(10^{-2}) - \mathcal{O}(10^{-1})$ where $Sr =\frac{fc}{u_\infty}$ is the non-dimensional frequency denoted as Strouhal number. The spatial oscillation of the shock wave leads to periodically alternating forces on the wing structure. The structural response to those dynamic loads, generally referred to as buffeting, have detrimental effects on the structural integrity and overall flight safety. Therefore, it defines a strict boundary to an aircraft's flight envelope. Economical and ecological challenges require aviation to make ever more efficient use of the available resources. A thorough knowledge of the aeroelastic behavior of aircraft is crucial in the endeavor to further reduce structural weight and optimize fuel consumption. Therefore, a comprehensive understanding of the buffet phenomenon is substantially important in the design process of future aircraft. 

%% Buffet has been researched for many years ... no thourough therz exists.
Although the buffet phenomenon has been intensively investigated for decades a comprehensive description of the underlying mechanisms is still subject to debate. The most widely accepted model explaining the self-sustained nature of the low-frequency shock oscillation has been introduced by Lee~\cite{lee2001}. The proposed mechanism is based on a closed feedback loop between the shock wave and the trailing edge flow. The interaction of the incoming boundary layer with the shock wave leads to the generation of vortices which convect downstream and eventually pass over the trailing edge. The transition from a wall-bounded flow to a free-shear layer causes the generation of pressure waves which propagate back upstream. The pressure waves interact with the shock wave and push the shock upstream leading to a thickening and ultimately a separation of the boundary layer downstream of the shock. The separation of the boundary layer mitigates or massively reduces the local wall-shear stress near the trailing edge flow. Thus, also the generation of pressure waves at the trailing edge is attenuated due to the decreased Lamb vector, i.e., the outer product of vorticity and the velocity vectors, and the shock moves back downstream closing the feedback loop. Good agreement with the buffet model proposed by Lee was found, among others, by Deck~\cite{deck2005}, Xiao et al.~\cite{xiao2006}, Hartmann et al.~\cite{hartmann2012, hartmann2013a, hartmann2013b} and Feldhusen-Hoffmann et al.~\cite{feldhusen2017, feldhusen2021}.
Lee's model was further developed by Hartmann et al.~\cite{hartmann2013b}, who proposed that the relevant interaction region of the shock wave and the pressure waves originating from the trailing edge is at the upper part of the shock wave instead of the shock foot. This formulation led to a significantly better agreement between the predicted and the measured buffet frequency.

In the pursuit to design ever more efficient aircraft, engine development is clearly moving toward larger and larger bypass ratios, so called ultra-high bypass ratio (UHBR) turbofan engines. The large diameter of such engines requires them to be mounted closely under the wing to ensure sufficient ground clearance without having to massively redesign the landing gear much more generously. As a result, the flow around the wing is significantly affected by the engine flow. In transonic flight, shock waves with all their detrimental effects, i.e., shock-induced separation, shock unsteadiness, and total pressure loss, can occur on the engine nacelle~\cite{dietz2008, spinner2021, spinner2022}. The buffet phenomenon is, however, highly sensitive to the upstream flow conditions, which are strongly altered by an upstream shock wave. Furthermore, the shear layer developing on the engine nacelle merges into the boundary layer on the pressure side of the airfoil. As a result, also  the flow in the vicinity of the trailing edge is influenced by the integration of UHBR engines. Since the trailing edge flow is a central element to trigger shock wave oscillations~\cite{lee2001, hartmann2013b} the quality of the buffet phenomenon can be significantly affected. 

In recent years, the buffet phenomenon has been studied extensively ~\cite{lee2001, xiao2006, deck2005, hartmann2012, hartmann2013a, hartmann2013b, feldhusen2017, feldhusen2021, giannelis2017, jacquin2009}. Yet, little is known about the interaction of engine induced flow disturbances and the involved dynamics. A recent experimental study by Spinner \& Rudnik \cite{spinner2022} investigates the phenomenon of shock buffet on the lower surface of the wing, which occurs at negative angles of attack due to the integration of the UHBR. To the best of the authors' knowledge, studies on the impact of engine nacelle integration on the buffet phenomenon on the upper wing surface do not exist in the archival literature. Yet, engine nacelle integration is an important aspect of the aerodynamic design of commercial aircraft. This means, it is of great interest to complement the current knowledge by the effect of engine induced flow disturbances on the buffet phenomenon.

In this paper, first studies are made to better understand the interaction between engine-induced disturbances of the flow field and the buffet phenomenon on the suction side of the wing. Using wall-modeled large-eddy simulations (WM-LES) the flow field around the OAT15A airfoil with a generic UHBR engine nacelle geometry and a baseline configuration without the nacelle is computed at buffet conditions. To allow the analysis of the upstream shock and the free-shear layer interacting with the airfoil boundary layer at reasonable computational cost the nacelle is modeled as a 2D-periodic flow-through nacelle. The flow field is analyzed focusing on the disturbance of the flow field due to the integration of the nacelle geometry. The resulting changes of the occurring shock dynamics are examined using the sparsity-promoting dynamic mode decomposition (SP-DMD).

The paper is organized as follows. In section~\ref{sec:methods}, the simulation framework and the numerical methods are briefly presented. In particular, the wall-modeling approach is discussed in subsection~\ref{ssec:wallmodel} and the SP-DMD algorithm is introduced in subsection~\ref{ssec:dmd}. The computational setup including the numerical mesh and the mesh refinement strategy is discussed in section~\ref{sec:setup}. The results are presented in section~\ref{sec:results} with subsection~\ref{ssec:flowfield} focusing on direct insights from the WM-LES. The results obtained performing SP-DMD on the simulation data are discussed in subsection~\ref{ssec:dmdanalysis}. Finally, conclusions are drawn in section~\ref{sec:conclusion}.

\section{Numerical Methods}\label{sec:methods}
All simulations are performed using the Cartesian solver of the m-AIA framework which was formerly denoted zonal flow solver (ZFS)~\cite{lintermann2020}. The governing equations are the compressible, unsteady Navier-Stokes equations. The equations are spatially filtered and solved on an unstructured, hierarchical Cartesian mesh using the finite-volume method. The advective upstream splitting method (AUSM) is used for the convective fluxes. A central difference scheme is employed for the discretization of the viscous fluxes. For the temporal integration, a second-order accurate 5-stage Runge-Kutta method is used. Small scale turbulence is accounted for by the approach of monotone integrated LES (MILES)~\cite{boris1992} such that no additional turbulence model is necessary~\cite{meinke2002}. Boundaries of embedded bodies are realized using a cut-cell method~\cite{schneiders2016}. Further details on the numerical methods used by the m-AIA simulation framework are given in~\cite{meinke2002, schneiders2016}.

\subsection{Wall-stress model}
\label{ssec:wallmodel}
\begin{figure}[htb]
\centering
\begin{tikzpicture}[scale=1.0]
    \path [use as bounding box] (0.0,0.0) rectangle (8.0, 4.75);
    \coordinate (S) at (0.0, 1.0);
    \coordinate (E) at (8.0, 0.0);
    \coordinate (VE) at (1.9, 3.375);
    \coordinate (VS) at (1.9, 0.5);
    \coordinate (VM) at (3.4, 3.375);
    \draw[step = 0.75, thin, lightgray] (0.01, 0.01) grid (8.0, 4.75);
    \draw[name path = vert, black, thick] (VS) -- (VE);
    \draw[fill, white] (S) to [bend left = 10] (E);
    \draw[name path = body, black] (S) to[bend left = 10] (E);
    \draw[fill, white] (0,0) -- (S) -- (E);
    \path[name intersections={of=body and vert,by=VS}];
    \draw[black, thick, name path = profile] (VS) to[out=0 , in=-100] (VM);

    \foreach \i in {1,...,4} {
        \coordinate (AI) at ($(VS) + ($0.25*\i*($(VE) - (VS)$)$)$);
        \path[name path = hor] (AI) -- ($(AI) + (10.0, 0.0)$);
        \path[name intersections={of=hor and profile, by=AE}];
        \draw[->, black, very thick] (AI) -- (AE);
    }

    \def \offset {0.4};
    \draw[decoration={brace, amplitude = 5pt}, decorate, very thick] ($(VS)-(\offset, 0.0)$) -- node[left=8pt, rectangle, draw = black, fill=white, rounded corners = 0.1cm, inner sep = 0.2cm] {$h_{wm}$} ($(VE)-(\offset, 0.0)$);
    
    \node[draw, circle, fill=white] (sp) at (AI){$\bm u$};
    \node[draw, circle, fill=white] (sf) at (VS){$\bm \tau$};
    %\node[draw, rounded corners = 0.2cm, fill=white] at (6,3.5) {\Large};
    \node[draw, rounded corners = 0.2cm, fill=white, inner sep = 0.4cm] at (6,2.5) (wm) {\large Wall Model};
    \draw[->, very thick] (sp) to[out=45 , in=90] (wm);
    \draw[->, very thick] (wm) to[out= -90, in=-45] (sf);
\end{tikzpicture}
\caption{Graphic representation of the wall-model mechanism, by courtesy of~\cite{luerkens2022}.}
\label{fig:wallmnodel}
\end{figure}
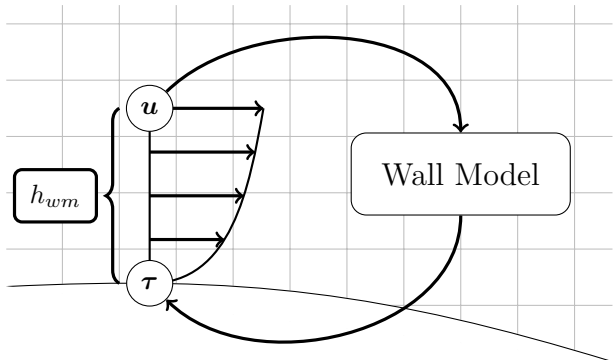
Traditional wall-resolved LES of boundary layer flows are restricted by the resolution of the viscous sublayer which requires extremely small cells at high Reynolds numbers. This results in a large amount of mesh cells and thus in a enormous demands on the available computer hardware that often can not be fulfilled. At the same time, the computational time step is restricted by the CFL condition leading to a very slow progress of the simulation in terms of convective time units $CTU = c/u_{\infty}$ making wall-resolved LES of high Reynolds number flows computationally very expensive.
Wall-modeling approaches are a convenient method to circumvent the massive constraints on mesh resolution for traditional LES of boundary layer flows while maintaining a high spatial and temporal resolution. Omitting the resolution of the viscous sublayer and only resolving large scale structures in the boundary layer WM-LES enables not only a significant reduction in the total number of cells but does also allow a substantially larger computational time step. Thereby, LES of dynamic phenomena at low Strouhal numbers $Sr = \frac{f\cdot c}{u_\infty}$ such as the buffet phenomenon can be performed even at high Reynolds numbers with reasonable computational effort. The applicability of such a wall-stress model to buffet flows has been demonstrated by Fukushima \& Kawai~\cite{fukushima2018}. 

In this study, an analytical wall-stress model is used to compute the wall-shear stress based on flow information obtained from the outer boundary layer. The function is an implicit single equation expression for the law of the wall introduced by Spalding~\cite{spalding1961}:
\begin{equation}\label{eq:spalding}
    y^+ = u^+ + e^{-\kappa B} \left\{ e^{\kappa u^+} - \sum_{n=0}^{4} \frac{(\kappa u^+)^n}{n!} \right\}
\end{equation}
with $u^+ = \frac{u_{\|}}{u_\tau}$, $y^+=\frac{h_{wm}u_\tau}{\nu}$, and $u_\tau = \sqrt{\tau_w / \rho_w}$. The subscript $(\bullet)_{\|}$ denotes projection onto the wall-tangential plane. The von Kármán constant $\kappa$ is set to $0.4$ and $B$ is $5.0$. Equation~\ref{eq:spalding} is evaluated at a sampling distance $y=h_{wm}$ normal to the surface which is illustrated in figure~\ref{fig:wallmnodel}. The implicit expression is solved iteratively for $u_\tau$ using Newton's method. To apply the  wall-shear stress to the boundary surface an additional correction loop within the viscous flux computation is performed adding an artificial viscosity $\mu_{wm}$ such that the wall-shear stress given by
\begin{equation}
  \left( \mu + \mu_{wm} \right) \frac{\partial u_{\|}}{\partial n} = \tau_{wm}
\end{equation}
is imposed. To prevent an obvious misapplication of the model in regions of separated flow the velocity gradient at the wall is checked for $\partial u_{\|}/\partial{n} < 0$. If flow separation is detected, the application of the wall-stress model is omitted by setting $\mu_{wm} = 0$.

\subsection{Dynamic mode decomposition}
\label{ssec:dmd}
The dynamic mode decomposition~\cite{schmid2010} is a data-driven technique that allows the decomposition of a given data field $\bm{q}(\bm{x},t)$ into a series representation of spatio-temporal modes $\phi_n$. Every mode is characterized by an individual, complex frequency $\lambda_n$, and amplitude $a_n$. The resulting series representation of the data field reads
\begin{align}
    \bm{q}(\bm{x},t) = \sum_n a_n e^{\lambda_n t} \bm{\phi}_n(\bm{x}).
\end{align}
For these modes characteristic dynamic phenomena of distinct frequencies can be determined and analyzed in detail. The buffet phenomenon is typically restricted to narrow frequency bands such that DMD can be an effective tool for the in-depth analysis of the isolated dynamics and the underlying physical mechanisms, which is reflected by its frequent use in the literature~\cite{kou2017, kou2018, gao2017, poplingher2019, feldhusen2021}. In general, a DMD performed on a set of $N$ snapshots results in $N-1$ modes, which are given as complex conjugate pairs. For a large number of snapshots this results in an inconveniently large number of modes to assess. The sparsity-promoting DMD algorithm proposed by Jovanović et al. \cite{jovanovic2014} addresses this problem by introducing a penalty term $\gamma\sum_{n=1}^{N-1}\lvert a_n\rvert$ into the minimization problem determining the modes, which reads
\begin{align}
\label{eq:spdmd}
    \underset{a}{\text{minimize}} \, \Vert \bm{V}^{N-1}_1 - \bm{\Phi}\bm{D}_a \bm{V}_{and} \Vert_F^2 + \gamma \sum_{n=1}^{N-1} \lvert a_n \rvert \,. 
\end{align}
Here, $\bm{V}_1^{N-1}$ is the discrete snapshot sequence and $\bm{\Phi}$ contains the dynamic modes given in matrix form by
\begin{align}
   \bm{V}_1^{N-1} = \left[ \bm{v}_1, \bm{v}_2, \ldots, \bm{v}_{N-1}\right], \quad \bm{\Phi} = \left[ \bm{\phi}_1, \bm{\phi}_2, \ldots, \bm{\phi}_{N-1}\right]\, .
\end{align}
The matrix $\bm{D}_a = \text{diag}(a)$ is the diagonal matrix of all optimized amplitudes $a_n$. The Vandermonde matrix $\bm{V}_{and}$ contains the so-called Ritz eigenvalues $\mu_n = e^{\lambda_n \Delta t}$ and reads
\begin{align}
    \bm{V}_{and} = \left(
    \begin{matrix}
    1 & \mu_1 & \ldots & \mu_1^{N-1} \\
    1 & \mu_2 & & \mu_2^{N_1} \\
    \vdots & \vdots & & \vdots \\
    1 & \mu_{N-1} & & \mu_{N-1}^{N-1}
    \end{matrix}
    \right).
\end{align}
When solving the minimization problem given by expression \ref{eq:spdmd}, increasing values of the penalty parameter $\gamma$ imply a stronger penalization of non-zero amplitudes, and thus result in an overall lower number of non-zero amplitude modes. More details on the SP-DMD algorithm are given in \cite{jovanovic2014}. The application of SP-DMD to transonic buffet is thoroughly discussed in \cite{feldhusen2021}.
\section{Computational Setup}
\label{sec:setup}
\begin{figure}
    \begin{subfigure}[t]{0.49\textwidth}
    \vskip 0pt
        \includegraphics[width=1\linewidth]{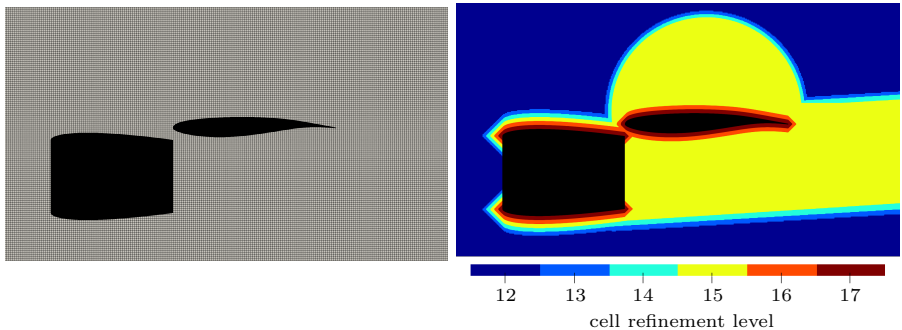}
        \vspace{16.5pt}
        \caption{Computational mesh of the nacelle configuration on the coarsest level.}
        \label{fig:grid_raw}
    \end{subfigure}
    \begin{subfigure}[t]{0.49\textwidth}
    \vskip 0pt
        \begin{overpic}[width=1\linewidth]{pictures/gridlevels_jet.png}
        \put(86.5,7){\footnotesize $17$}
        \put(71,7){\footnotesize $16$}
        \put(55.5,7){\footnotesize $15$}
        \put(39.7,7){\footnotesize $14$}
        \put(24.4,7){\footnotesize $13$}
        \put(8.7,7){\footnotesize $12$}
        \put(30, 1){\footnotesize $\text{cell refinement level}$}
        \end{overpic}
        \caption{Regions of grid refinement around the OAT15A airfoil and the nacelle.}
        \label{fig:grid_ref}
    \end{subfigure}
    \caption{Overview of the computational mesh of the nacelle configuration, by courtesy of~\cite{luerkens2022}.}
\end{figure}
In the following, the computational setup of the nacelle configuration is presented. Note that except for the integration of the nacelle geometry 
the baseline configuration, i.e., no engine is taken into account, and the UHBR-airfoil configuration, and the corresponding mesh refinement both setups are identical and therefore no further differentiation is made unless otherwise necessary. A setup with only mild flow separation while still providing low-frequency shock oscillations is considered. Based on the work by Fukushima \& Kawai~\cite{fukushima2018}, Jacquin et al.~\cite{jacquin2009} and Deck~\cite{deck2005} the angle of attack was set to $\alpha = 3.5^\circ$ and the freestream Mach number to $Ma_\infty = 0.73$. In agreement with the associated experimental campaign by Schauerte \& Schreyer~\cite{schauerte2022} the chord based Reynolds number was set to $Re_c = 2\cdot10^6$. 

For the nacelle, a generic configuration based on a NACA 0012 airfoil was generated. The outer geometry of the nacelle geometry with respect to the local chord length is roughly based on the UHBR flow-through nacelle by Spinner \& Rudnik~\cite{spinner2021}, which was designed for the Airbus XRF-1 research model. That is, the inlet and the outlet diameter were set to ${d_{nac} = 0.45c}$ and the nacelle length to ${l_{nac} = 0.75c}$.

The grid dimensions in the x-, y-, and z-coordinate directions $L_x \times L_y \times L_z = 25.4c \times 24.9c \times 0.05c$, which proved to be sufficient for comparable flows by Zauner \& Sandham~\cite{zauner2020} and Moise et al.~\cite{moise2021}. The grid is locally refined within the boundary layer satisfying the mesh requirements for WM-LES proposed by Kawai \& Larsson~\cite{kawai2012}, i.e., at least 20 cells are located within the boundary layer at $x_{oat}/c = 0.2$ and $x_{nac}/c = 0.2$. Since the acoustics in the vicinity of the trailing edge is a key element of established buffet models additional a-priori refinement is applied on the suction side of the airfoil and in the wake region of the airfoil and the nacelle. The Cartesian mesh of the nacelle configuration on the coarsest refinement level is shown in figure~\ref{fig:grid_raw}. Regions of additional mesh refinement are highlighted by a color distribution in figure~\ref{fig:grid_ref}. The local cell length is given by
\begin{align}
    L_{lvl} = L_0\cdot\left( \frac{1}{2} \right)^{lvl}
\end{align} where $lvl$ is the cell refinement level and $L_0$ is the mesh base length, which in this case corresponds to the longest mesh dimension of $L_x = 25.4c$.
The resulting number of Cartesian cells for both configurations is given in table~\ref{tab:meshsize}.
{\def\arraystretch{1.5}\tabcolsep=5pt
\begin{table}[htb]
    \centering
    \begin{tabular}{|c|c|c|}
    \hline
      & baseline & with nacelle \\
     \hline
      number of cells & $0.73\cdot 10^9$ & $1.24\cdot 10^9$ \\
      \hline
    \end{tabular}
    \caption{Mesh size of the baseline configuration without nacelle and the configuration including the nacelle.}
    \label{tab:meshsize}
\end{table}
}

The wall model is applied to all solid wall boundaries. Following the requirements proposed by Kawai \& Larsson~\cite{kawai2012} the evaluation distance $h_{wm}$ of the wall model is set to a distance of approximately three cell lengths from the wall. A characteristic outflow boundary condition is used for the main outflow plane to suppress spurious reflections. Standard in- and outflow boundary conditions with sponge layers are applied to the remaining in- and outflow planes in the x- and y-direction. Periodicity is applied to spanwise boundaries, which is a great simplification of the in general highly three-dimensional flow field around the engine nacelle. This, however, allows the study of the effect of the upstream shock wave and the free-shear layer introduced by the nacelle geometry onto the buffet phenomenon while keeping the total number of grid points at a manageable level. At the nacelle inlet, a standard outflow boundary condition is applied. For the engine outlet, a $1/7$th power law velocity profile is prescribed while maintaining mass conservation with the engine inlet. Hence, the setup represents a simple flow-through nacelle.

To enforce a controlled transition of the boundary layer into a turbulent state the tripping method by Schlatter and \"Orl\"u~\cite{schlatter2012} is employed at $x/c = 0.07$ on both sides of the airfoil.

\section{Results}\label{sec:results}
For both setups, data were collected for about 70 convective time units $CTU~=~c/u_\infty$. Spanwise averaged surface data of the OAT15A airfoil and the nacelle were sampled at a sampling interval of $4.1\cdot10^{-3}\,CTU$ at every $1\%$ of the airfoil's chord length. Additionally, wall normal velocity and pressure profiles were determined at every $5\%$ of the airfoil's chord length at the same sampling interval of $4.1\cdot10^{-3}\,CTU$. For the analysis of the shock dynamics with DMD, volumetric data of the flow field were collected at a sampling interval of $0.1025\,CTU$.

\subsection{General flow field}\label{ssec:flowfield}
\begin{figure}[htb]
    \centering
    \includegraphics[width=\textwidth]{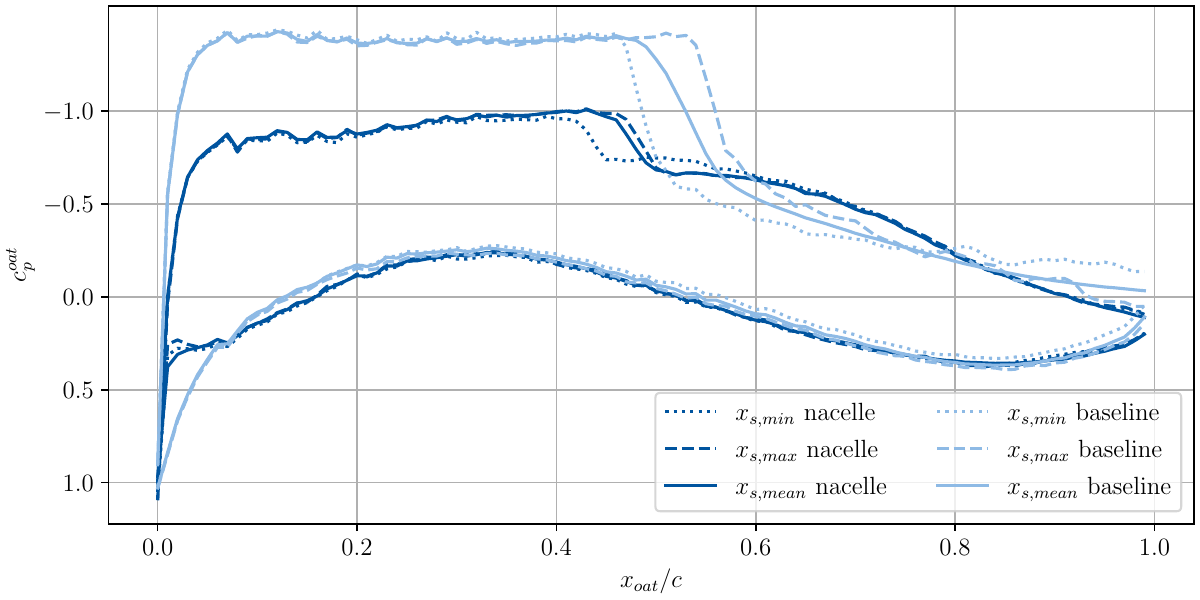}
    \caption{Mean pressure coefficient and instantaneous pressure coefficient on the airfoil at the most upstream and downstream shock position of both configurations.}
    \label{fig:cp}
\end{figure}

\begin{figure}[htb]
    \centering
    \includegraphics[width=\textwidth]{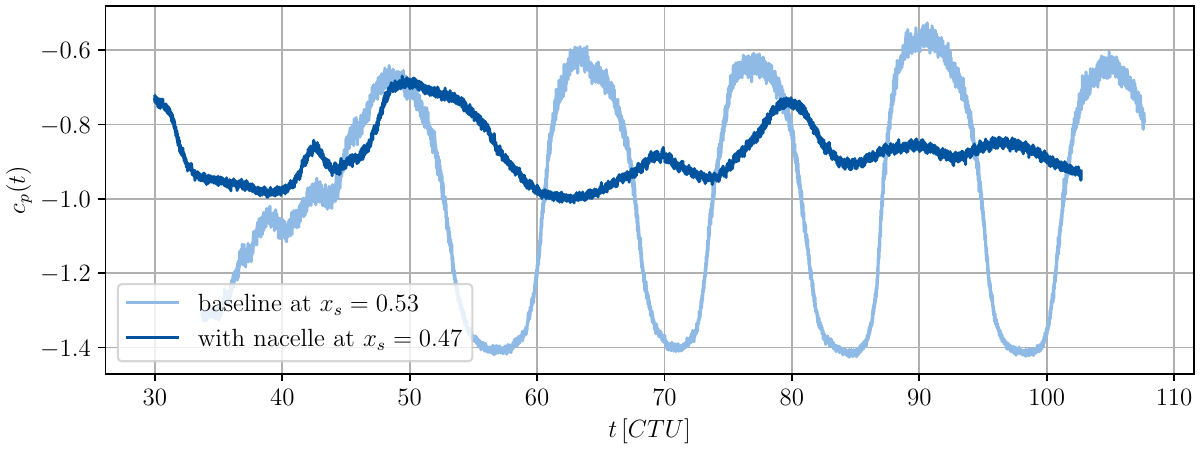}
    \caption{Time history of the surface pressure at the mean shock position for both configurations.}
    \label{fig:psignal}
\end{figure}

\begin{figure}[htb]
    \centering
    \begin{subfigure}[t]{0.49\textwidth}
    \includegraphics[width=\linewidth]{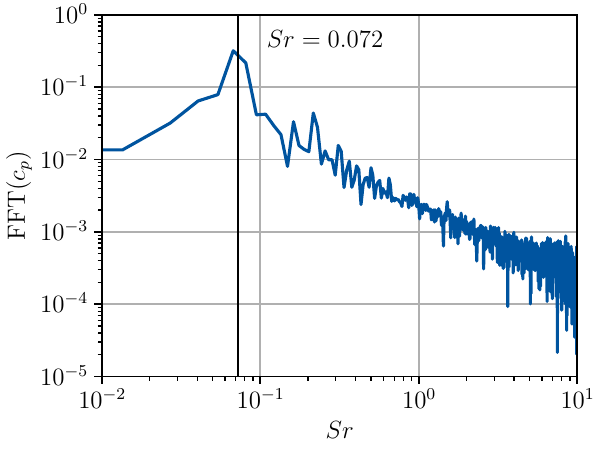}
    \caption{baseline: $x_{s,mean} = 0.53c$}
    \label{fig:cpnj_fft}
    \end{subfigure}
    \begin{subfigure}[t]{0.49\textwidth}
    \includegraphics[width=\linewidth]{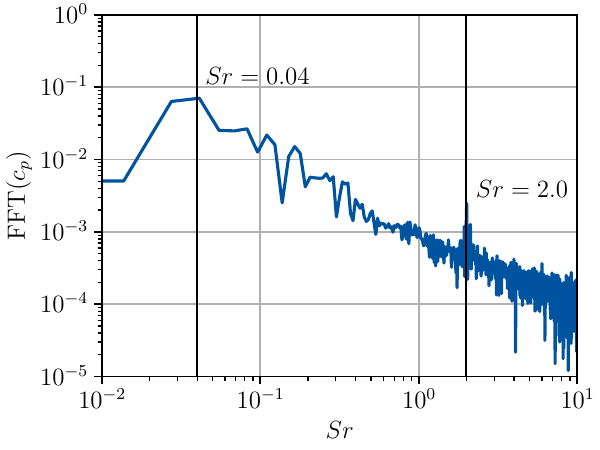}
    \caption{with nacelle $x_{s,mean} = 0.47c$}
    \label{fig:cpj_fft}
    \end{subfigure}
    \caption{FFT of the pressure signal at the mean shock position for both configurations.}
\end{figure}

\begin{figure}[htb]
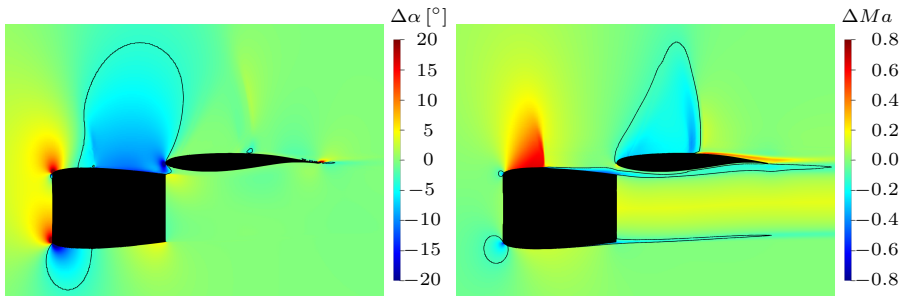

    \centering
    \begin{subfigure}[t]{0.49\textwidth}
     \begin{overpic}[width=\linewidth]{pictures/dAlpha.png}
     \put(90,2.4){\footnotesize $-20$}
     \put(90,9.3){\footnotesize $-15$}
     \put(90,16.2){\footnotesize $-10$}
     \put(91,23.0){\footnotesize $-5$}
     \put(94.5,29.8){\footnotesize $0$}
     \put(94.5,36.6){\footnotesize $5$}
     \put(93,43.4){\footnotesize $10$}
     \put(93,50.3){\footnotesize $15$}
     \put(93,57.2){\footnotesize $20$}
     \put(87,62.5){\footnotesize $\Delta\alpha\,[^\circ]$}
     \end{overpic}
    \caption{Flow angle deviation $\Delta\alpha$. Black lines show iso-contours at $\Delta\alpha = -3.5^\circ.$}
    \label{fig:diff_alpha}
    \end{subfigure}
    \begin{subfigure}[t]{0.49\textwidth}
    \begin{overpic}[width=\linewidth]{pictures/dMa.png}
     \put(90,2.4){\footnotesize $-0.8$}
     \put(90,9.3){\footnotesize $-0.6$}
     \put(90,16.2){\footnotesize $-0.4$}
     \put(90,23.0){\footnotesize $-0.2$}
     \put(93.8,29.8){\footnotesize $0.0$}
     \put(93.8,36.6){\footnotesize $0.2$}
     \put(93.8,43.4){\footnotesize $0.4$}
     \put(93.8,50.3){\footnotesize $0.6$}
     \put(93.8,57.2){\footnotesize $0.8$}
     \put(87,62.5){\footnotesize $\Delta Ma$}
    \end{overpic}
    \caption{Mach number deviation $\Delta Ma$. Black lines show iso-contours at $\Delta Ma=-0.1$.}
    \label{fig:diff_mach}
    \end{subfigure}
    \caption{Deviation of the flow field of the nacelle configuration from the baseline configuration.}
\end{figure}

In figure~\ref{fig:cp}, the airfoil's mean and instantaneous pressure coefficients at the most upstream and most downstream shock position are shown for both configurations. It is remarkable that the modeling of the engince nacelle as a flow-through nacelle affects the pressure side only marginally. It is, however, obvious that the integration of the nacelle leads to a significantly weaker shock and an upstream displacement of the mean shock location on the suction side. The mean shock position is determined at approximately $x_s = 0.53$ for the baseline configuration and $x_s = 0.47$ for the configuration including the nacelle. Considering the most upstream and most downstream shock locations, the nacelle configuration exhibits a noticeably lower amplitude in the shock dynamics compared to the baseline configuration. This finding is supported by the time-resolved pressure signal at the mean shock position shown in figure~\ref{fig:psignal}. After a transient phase the baseline configuration is characterized by a stable -- almost harmonic -- oscillation. The pressure signal of the nacelle configuration, however, exhibits highly variable dynamics of noticeably lower amplitude. The frequency spectra of these pressure signals are shown in figures~\ref{fig:cpnj_fft} and \ref{fig:cpj_fft}. The consistent low-frequency shock oscillation of the baseline configuration is evidenced in a clear peak at $Sr= 0.072$. The spectrum of the nacelle configuration, however, is characterized by a much more broadband peak in the range of $0.03 < Sr < 0.04$ with the expected lower amplitude. An additional peak at $Sr = 2.0$ is prominent, which is not observed in the baseline configuration.
To understand this alteration to the shock dynamics, the disturbance of the flow onto the airfoil due to the integration of the engine nacelle will be considered next. In figure~\ref{fig:diff_alpha} the deviation of the local flow angle of the nacelle configuration from the baseline configuration given by 
\begin{align}
 \Delta \alpha = \arctan \left( \frac{\bar{u}}{\bar{v}}  \right)_{nac} - \arctan \left( \frac{\bar{u}}{\bar{v}}  \right)_{base}
\end{align}
is shown. It is evident that immediately upstream of the airfoil the flow is deflected to smaller angles of attack by the contour of the engine nacelle. In figure~\ref{fig:diff_mach} the deviation of the local Mach number of the nacelle configuration from the baseline configuration defined by 
\begin{align}
    \Delta Ma = \left( \frac{\lvert \bm{\bar{u}} \rvert}{\sqrt{\gamma \bar{p}/\bar{\rho}}} \right)_{nac} - \left(\frac{\lvert \bm{\bar{u}} \rvert}{\sqrt{\gamma \bar{p}/\bar{\rho}}} \right)_{base}
\end{align}
is shown. As a consequence of the shock and the boundary layer on the upper part of the nacelle the Mach number of the flow onto the airfoil is significantly reduced. 

The buffet phenomenon is, however, highly sensitive to the angle of attack and the Mach number. Considering the numerical and experimental studies on the OAT15A airfoil at comparable conditions \cite{deck2005, jacquin2009, schauerte2022}, it is reasonable that the reduction of the angle of attack and the Mach number due to the integration of the engine nacelle results in a shift of the dynamics from fully developed buffet to pre-developed buffet.
Furthermore, the Mach number in the free-shear layer downstream of the trailing edge is evidently affected by the integration of the nacelle. The wake flow coming from the upper part of the nacelle interacts with the boundary layer on the pressure side of the airfoil leading to a reduced local Mach number. On the suction side, the weaker shock wave leads to a reduced flow  deceleration and to a weaker boundary layer separation and thus, to a higher local Mach number. 

\begin{figure}[htb]
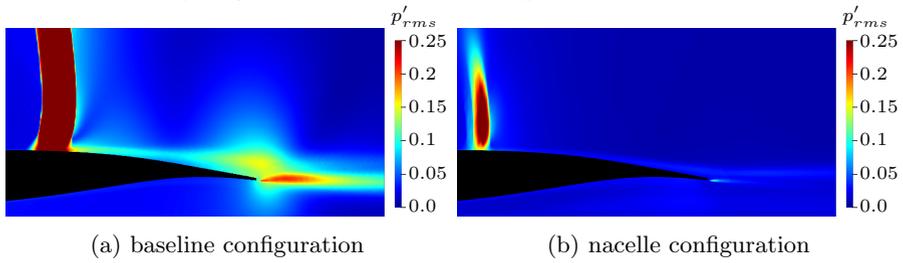

    \centering
    \begin{subfigure}[t]{0.49\textwidth}
     \begin{overpic}[width=\linewidth]{pictures/pfluc_nj.png}
     \put(91, 1.0){\footnotesize $0.0$}
     \put(91, 8.5){\footnotesize $0.05$}
     \put(91, 16.2){\footnotesize $0.1$}
     \put(91, 23.8){\footnotesize $0.15$}
     \put(91, 31.3){\footnotesize $0.2$}
     \put(91, 38.8){\footnotesize $0.25$}
     \put(87, 44.5){\footnotesize $p'_{rms}$}
     \end{overpic}
    \caption{baseline configuration}
    \label{fig:prms_nj}
    \end{subfigure}
    \begin{subfigure}[t]{0.49\textwidth}
    \begin{overpic}[width=\linewidth]{pictures/pfluc_j.png}
     \put(91, 1.0){\footnotesize $0.0$}
     \put(91, 8.5){\footnotesize $0.05$}
     \put(91, 16.2){\footnotesize $0.1$}
     \put(91, 23.8){\footnotesize $0.15$}
     \put(91, 31.3){\footnotesize $0.2$}
     \put(91, 38.8){\footnotesize $0.25$}
     \put(87, 44.5){\footnotesize $p'_{rms}$}
    \end{overpic}
    \caption{nacelle configuration}
    \label{fig:prms_j}
    \end{subfigure}
    \caption{Contours of $p'_{rms}$ of the baseline configuration and the nacelle configuration.}
\end{figure}

Contours of $p'_{rms}$ of both configurations are depicted in figures~\ref{fig:prms_nj} and \ref{fig:prms_j}. The more upstream mean shock position and the lower amplitude of the shock oscillation indicated by figure \ref{fig:cp} are confirmed. Additionally, the pressure fluctuations of the nacelle configuration are massively reduced. Therefore, the integration of the engine nacelle can not merely be considered as a perturbation of the global flow topology, but also has a notable impact on the acoustics between trailing edge and shock wave. The shock-induced separation on the suction side and the noise generated by the trailing edge shear flow, however, are essential to established buffet models \cite{lee2001, hartmann2013b}. This means the alteration of the shock dynamics due to the integration of the nacelle does have an important impact on the overall aerodynamics of the system. 

\subsection{DMD analysis}\label{ssec:dmdanalysis}
\begin{figure}[htb]
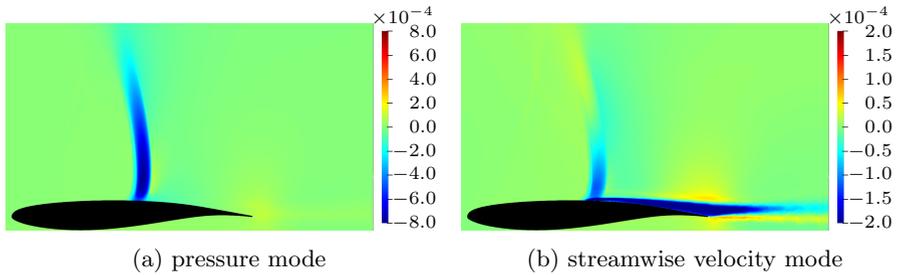

    \centering
    \begin{subfigure}[t]{0.495\textwidth}
    \begin{overpic}[width=\linewidth]{pictures/dmd_nj_p_0072.png}
    \put(86.5, 0.6){\footnotesize $-8.0$}
    \put(86.5, 6.0){\footnotesize $-6.0$}
    \put(86.5, 11.5){\footnotesize $-4.0$}
    \put(86.5, 16.75){\footnotesize $-2.0$}
    \put(90.3, 22.1){\footnotesize $0.0$}
    \put(90.3, 27.5){\footnotesize $2.0$}
    \put(90.3, 32.8){\footnotesize $4.0$}
    \put(90.3, 38.2){\footnotesize $6.0$}
    \put(90.3, 43.7){\footnotesize $8.0$}
    \put(81.8, 46.7){\footnotesize $\times 10^{-4}$}
    \end{overpic}
    \caption{pressure mode}
    \label{fig:dmd_nj_p_buffet}
    \end{subfigure}
    \begin{subfigure}[t]{0.495\textwidth}
    \begin{overpic}[width=\linewidth]{pictures/dmd_nj_u_0072.png}
    \put(86.5, 0.6){\footnotesize $-2.0$}
    \put(86.5, 6.0){\footnotesize $-1.5$}
    \put(86.5, 11.5){\footnotesize $-1.0$}
    \put(86.5, 16.75){\footnotesize $-0.5$}
    \put(90.3, 22.1){\footnotesize $0.0$}
    \put(90.3, 27.5){\footnotesize $0.5$}
    \put(90.3, 32.8){\footnotesize $1.0$}
    \put(90.3, 38.2){\footnotesize $1.5$}
    \put(90.3, 43.7){\footnotesize $2.0$}
    \put(81.8, 46.7){\footnotesize $\times 10^{-4}$}
    \end{overpic}
    \caption{streamwise velocity mode}
    \label{fig:dmd_nj_u_buffet}
    \end{subfigure}    
    \caption{Buffet modes at $Sr=0.072$ of the pressure and the streamwise velocity fields of the baseline configuration.}
\end{figure}
\begin{figure}[htb]
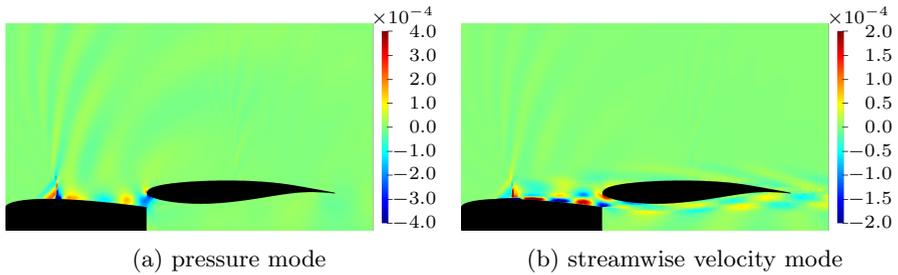

    \centering
    \begin{subfigure}[t]{0.495\textwidth}
    \begin{overpic}[width=\linewidth]{pictures/dmd_j_p_201.png}
    \put(86.5, 0.6){\footnotesize $-4.0$}
    \put(86.5, 6.0){\footnotesize $-3.0$}
    \put(86.5, 11.5){\footnotesize $-2.0$}
    \put(86.5, 16.75){\footnotesize $-1.0$}
    \put(90.3, 22.1){\footnotesize $0.0$}
    \put(90.3, 27.5){\footnotesize $1.0$}
    \put(90.3, 32.8){\footnotesize $2.0$}
    \put(90.3, 38.2){\footnotesize $3.0$}
    \put(90.3, 43.7){\footnotesize $4.0$}
    \put(81.8, 46.7){\footnotesize $\times 10^{-4}$}
    \end{overpic}
    \caption{pressure mode}
    \label{fig:dmd_j_p_nac}
    \end{subfigure}
    \begin{subfigure}[t]{0.495\textwidth}
    \begin{overpic}[width=\linewidth]{pictures/dmd_j_u_201.png}
    \put(86.5, 0.6){\footnotesize $-2.0$}
    \put(86.5, 6.0){\footnotesize $-1.5$}
    \put(86.5, 11.5){\footnotesize $-1.0$}
    \put(86.5, 16.75){\footnotesize $-0.5$}
    \put(90.3, 22.1){\footnotesize $0.0$}
    \put(90.3, 27.5){\footnotesize $0.5$}
    \put(90.3, 32.8){\footnotesize $1.0$}
    \put(90.3, 38.2){\footnotesize $1.5$}
    \put(90.3, 43.7){\footnotesize $2.0$}
    \put(81.8, 46.7){\footnotesize $\times 10^{-4}$}
    \end{overpic}
    \caption{streamwise velocity mode}
    \label{fig:dmd_j_u_nac}
    \end{subfigure}    
    \caption{Modes of the shock-wave/boundary-layer interaction at $Sr=2.01$ of the pressure and the streamwise velocity fields of the nacelle configuration.}
\end{figure}
\begin{figure}[htb]
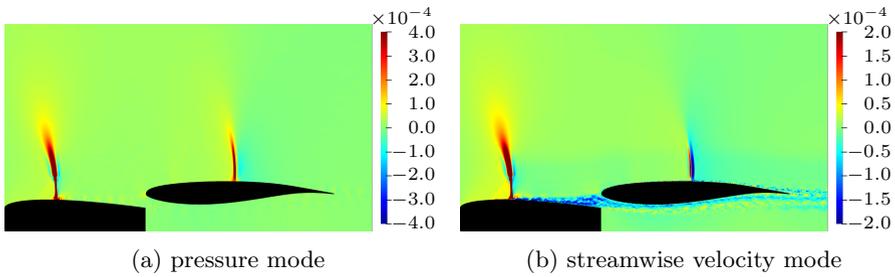

    \centering
    \begin{subfigure}[t]{0.495\textwidth}
    \begin{overpic}[width=\linewidth]{pictures/dmd_j_p_0043.png}
    \put(86.5, 0.6){\footnotesize $-4.0$}
    \put(86.5, 6.0){\footnotesize $-3.0$}
    \put(86.5, 11.5){\footnotesize $-2.0$}
    \put(86.5, 16.75){\footnotesize $-1.0$}
    \put(90.3, 22.1){\footnotesize $0.0$}
    \put(90.3, 27.5){\footnotesize $1.0$}
    \put(90.3, 32.8){\footnotesize $2.0$}
    \put(90.3, 38.2){\footnotesize $3.0$}
    \put(90.3, 43.7){\footnotesize $4.0$}
    \put(81.8, 46.7){\footnotesize $\times 10^{-4}$}
    \end{overpic}
    \caption{pressure mode}
    \label{fig:dmd_j_p_buffet}
    \end{subfigure}
    \begin{subfigure}[t]{0.495\textwidth}
    \begin{overpic}[width=\linewidth]{pictures/dmd_j_u_0047.png}
    \put(86.5, 0.6){\footnotesize $-2.0$}
    \put(86.5, 6.0){\footnotesize $-1.5$}
    \put(86.5, 11.5){\footnotesize $-1.0$}
    \put(86.5, 16.75){\footnotesize $-0.5$}
    \put(90.3, 22.1){\footnotesize $0.0$}
    \put(90.3, 27.5){\footnotesize $0.5$}
    \put(90.3, 32.8){\footnotesize $1.0$}
    \put(90.3, 38.2){\footnotesize $1.5$}
    \put(90.3, 43.7){\footnotesize $2.0$}
    \put(81.8, 46.7){\footnotesize $\times 10^{-4}$}
    \end{overpic}
    \caption{streamwise velocity mode}
    \label{fig:dmd_j_u_buffet}
    \end{subfigure}    
    \caption{Buffet modes of the pressure field at $Sr=0.043$ and the streamwise velocity field at $Sr=0.047$ of the nacelle configuration.}
\end{figure}
To further analyze the dynamics and investigate the underlying physical mechanisms with respect to established buffet models, SP-DMD was performed on data of the pressure and the streamwise velocity field. The penalty parameter $\gamma$ of the SP-DMD algorithm was varied to obtain only the most relevant modes. The performance loss defined by  
\begin{align}
    \Pi_{loss} = \frac{\|\bm{V}_1^N - \bm{\phi} \bm{D}_a\bm{V}_{and}\|_F}{\|\bm{V}_1^N\|_F} \cdot 100 \%
\end{align}
gives the relative deviation of the reconstructed field from the original snapshot sequence and allows the evaluation of the quality of the flow field reconstruction from the selected modes. More details on the SP-DMD algorithm are given in \cite{jovanovic2014, feldhusen2021}.

Performing SP-DMD on the pressure field data of the baseline configuration with $\gamma = 31257.44$ leaves only one relevant mode at a frequency of $Sr = 0.072$ which agrees well with the peak found in the FFT results of the surface pressure signal at the mean shock position given in figure \ref{fig:cpnj_fft}. Contours of the mode's amplitude are given in figure \ref{fig:dmd_nj_p_buffet}. They clearly show the association with the shock buffet. The performance loss at the given $\gamma$ is $1.7\%$ such that in return $98.3\%$ of the dynamics of the pressure field can be reconstructed using only this mode and its complex conjugate. Lowering the penalty parameter $\gamma$ merely introduces upper or lower harmonics of the buffet frequency into the DMD spectrum. From that it is obvious that the dynamics of the pressure field is solely dominated by the buffet mode. The corresponding mode of the streamwise velocity field at the same frequency of $Sr = 0.072$ is given in figure \ref{fig:dmd_nj_u_buffet}. Apart from the shock movement, this mode also reveals the periodic thickening and shrinking of the boundary layer past the shock wave, which is essential for established buffet models \cite{lee2001, hartmann2013b}. It is remarkable that there are two regions of changing sign in the vicinity of the trailing edge. These regions coincide with the peaks in the pressure fluctuations shown in figure \ref{fig:prms_nj}. The change of sign in the streamwise velocity mode implies a temporal variation of the local shear rate. Fluctuations in the local shear rate are, however, part of the driving acoustic source term in free-shear flows, which is the perturbed Lamb vector given by
\begin{align}
    \bm{L'} = (\bm{\omega} \times \bm{u})'
\end{align}
where $\bm{\omega}$ is the vorticity vector \cite{ewert2003}. 
One can conclude that the origin of the pressure fluctuations in the vicinity of the trailing edge is due to acoustic perturbations which agrees well with the established buffet models \cite{lee2001, hartmann2013b}. It should also be noted that both the pressure and velocity modes have a Ritz eigenvalue of close to $1$ such that the transient component of these modes is negligible.

Performing SP-DMD on the pressure field data of the nacelle configuration with $\gamma = 24770.98$ leaves a single remaining mode at a frequency of $Sr = 2.01$. A peak at this very frequency was already found in the FFT results in figure~\ref{fig:cpj_fft}. The real parts of the corresponding pressure and streamwise velocity modes are shown in figures \ref{fig:dmd_j_p_nac} and \ref{fig:dmd_j_u_nac}. It is obvious that this dynamic feature is restricted almost completely to the flow on the upper surface of the nacelle and can be associated with the shock-induced variation of the boundary layer on the nacelle and the subsequent vortex generation. The order of magnitude of the frequency generally agrees well with the findings of Moise et al. \cite{moise2022} who investigated the effect of boundary layer tripping on the buffet characteristics. The vortices passing over the trailing edge of the nacelle primarily convect along the pressure side of the airfoil and are subject to substantial dissipation. On the suction side, these structures are only poorly established. In particular, the shock wave on the airfoil suction side exhibits no considerable dynamic features and appears to be negligibly affected at this frequency. 

Lowering the penalty parameter to $\gamma = 12328.57$ includes the low-frequency mode at $Sr = 0.043$ which agrees well with the plateau-like low-frequency region in figure \ref{fig:cpj_fft}. The real part of the corresponding pressure mode is given in figure \ref{fig:dmd_j_p_buffet}. The corresponding streamwise velocity mode is detected at a slightly higher frequency of $Sr = 0.047$ and is shown in figure \ref{fig:dmd_j_u_buffet}. It is remarkable that both the nacelle and the airfoil shock display dynamic features at the respective frequency. The spatial amplitude of the shock motion, however, is clearly smaller compared to the buffet found for the baseline configuration. Considering the flow field disturbances by the nacelle geometry discussed in section \ref{ssec:flowfield}, this supports the suggestion that the nacelle suppresses developed buffet at the given flow parameters and a pre-developed state of buffet is observed. The shared dynamic features, however, suggest that a coupling mechanism between the dynamics of the nacelle shock and the airfoil shock exists. Note that similar to the baseline configuration, the streamwise velocity mode exhibits a periodic behavior of the nacelle boundary layer downstream of the shock. The wake flow of the nacelle merges into the boundary layer on the airfoil pressure side, such that eventually also the trailing edge flow of the airfoil is affected by the shock dynamics on the nacelle. Thereby, a connection between the dynamics of the nacelle shock and the mechanisms of existing buffet models \cite{lee2001, hartmann2013b} can be expected. Yet, as discussed in section \ref{ssec:flowfield} the acoustic field at the airfoil trailing edge is considerably weaker for the nacelle configuration such that determining the exact mechanism will require further scrutiny including detailed investigation of the acoustic source terms in the vicinity of the trailing edge.

% \begin{figure}[htb]
%     \centering
%     \begin{subfigure}[t]{\textwidth}
%     \includegraphics[width=\textwidth]{pictures/sparsity_nj.pdf}
%     \caption{without nacelle}
%     \label{fig:spar_nj}
%     \end{subfigure}
%     \begin{subfigure}[t]{\textwidth}
%     \includegraphics[width=\textwidth]{pictures/sparsity_j.pdf}
%     \caption{with nacelle}
%     \label{fig:spar_j}
%     \end{subfigure}
%     \caption{Results of the sparsity-promoting algorithm of the streamwise velocity-component of the baseline configuration (a) and the configuration with nacelle (b).}
% \end{figure}

% \begin{figure}[htb]
%     \centering
%     \begin{subfigure}[t]{\textwidth}
%     \includegraphics[width=\textwidth]{pictures/dmd_nj.pdf}
%     \caption{without nacelle}
%     \label{fig:dmd_nj}
%     \end{subfigure}
%     \begin{subfigure}[t]{\textwidth}
%     \includegraphics[width=\textwidth]{pictures/dmd_j.pdf}
%     \caption{with nacelle}
%     \label{fig:dmd_j}
%     \end{subfigure}    
%     \caption{DMD-spectrum of the streamwise velocity-component of the baseline configuration (a) and the configuration with nacelle(b). Modes selected by SP-DMD are shown in red.}
% \end{figure}

\section{Conclusion}\label{sec:conclusion}
WM-LES of the OAT15A complemented by a generic 2D nacelle geometry and a corresponding baseline configuration without the nacelle have been performed. The baseline case exhibits a well developed low-frequency shock oscillation at $Sr=0.072$. Investigation of the pressure fluctuations in the vicinity of the trailing edge and the flow field dynamics extracted by means of SP-DMD agree well with established buffet models. 
The nacelle configuration is characterized by a system of two shocks, one on the upper part of the engine nacelle and one on the airfoil suction side. Both shocks are subject to a low-frequency oscillation at the same frequency of $Sr=0.043$ which, however, has a smaller spatial amplitude than the buffet found for baseline configuration. The occurrence of a shared dynamic mode of the nacelle and the airfoil shock suggests the existence of a coupling mechanism, the exact determination of which requires further investigation. The analysis of the pressure field reveals a considerable attenuation of the pressure fluctuation in the vicinity of the trailing edge. 
Comparison of the mean flow fields of both configurations shows a significant disturbance of the flow conditions onto the airfoil. That is, both Mach number and local angle of attack are reduced by the introduction of the nacelle. The buffet phenomenon is, however, highly sensitive to variations of angle of attack and Mach number. Considering the mitigated and irregular shock dynamics, it is therefore stated that the altered flow conditions lead to a shift from fully developed buffet to pre-developed buffet. 

While the reduction of the UHBR nacelle to a 2D-periodic geometry obviously is a huge simplification it allowed the controlled study of the altered flow field and the occurring dynamics at reasonable computational cost. Even for such a generic case, the multitude of disturbances present in the flow makes the identification of a single cause for the altered shock dynamics and the coupling between the nacelle and the airfoil shock a challenging task that requires further investigation.  

%%%%%%%%%%%%%%%%%%%%%%%%%%%%%%%%%%%%%%%%%%%%%%%%%%%%%%%%%%%%%%%%%
%%                     BACKMATTER                              %%
%%%%%%%%%%%%%%%%%%%%%%%%%%%%%%%%%%%%%%%%%%%%%%%%%%%%%%%%%%%%%%%%%
\backmatter

%%%%%%%%%%%%%%%%%%%%%%%%%%%%%%%%%%%%%%%%%%%%%%%%%%%%%%%%%%%%%%%%%
%%            ADD ANY DECLARATION REGARDING FOR2895            %%
%%              OR PROVIDERS OF RESOURCES HERE                 %%
%%%%%%%%%%%%%%%%%%%%%%%%%%%%%%%%%%%%%%%%%%%%%%%%%%%%%%%%%%%%%%%%%

\bmhead{Acknowledgments} The authors gratefully acknowledge the Gauss Centre for Supercomputing e.V. (www.gauss-centre.eu) for funding this project by providing computing time on the GCS Supercomputer HAWK at Höchstleistungsrechenzentrum Stuttgart (www.hlrs.de).
\\
\\
The authors greatfully acknowledge the Deutsche Forschungsgemeinschaft DFG (German Research Foundation) for funding this work in the framework of the research unit FOR 2895. \\
\\
The authors also wish to thank ONERA for providing the OAT15A airfoil geometry.

%%%%%%%%%%%%%%%%%%%%%%%%%%%%%%%%%%%%%%%%%%%%%%%%%%%%%%%%%%%%%%%%%
%%            APPENDICES CAN BE ADDED HERE                     %%
%%%%%%%%%%%%%%%%%%%%%%%%%%%%%%%%%%%%%%%%%%%%%%%%%%%%%%%%%%%%%%%%%

\begin{appendices}

\end{appendices}

%%===========================================================================================%%
%% If you are submitting to one of the Nature Portfolio journals, using the eJP submission   %%
%% system, please include the references within the manuscript file itself. You may do this  %%
%% by copying the reference list from your .bbl file, paste it into the main manuscript .tex %%
%% file, and delete the associated \verb+\bibliography+ commands.                            %%
%%===========================================================================================%%
\newpage
\bibliography{sn-bibliography}% common bib file

\begin{thebibliography}{10}
\providecommand{\url}[1]{{#1}}
\providecommand{\urlprefix}{URL }
\providecommand{\doi}[1]{\url{https://doi.org/#1}}
\bibcommenthead

\bibitem{lee2001}
B.H.K. Lee, Self-sustained shock oscillations on airfoils at transonic speeds.
\newblock Progress in Aerospace Sciences \textbf{37}(2), 147--196 (2001).
\newblock \doi{10.1016/S0376-0421(01)00003-3}

\bibitem{deck2005}
S.~Deck, Numerical simulation of transonic buffet over a supercritical airfoil.
\newblock AIAA Journal \textbf{43}(7), 1556--1566 (2005).
\newblock \doi{10.2514/1.9885}

\bibitem{xiao2006}
Q.~Xiao, H.M. Tsai, F.~Liu, Numerical study of transonic buffet on a
  supercritical airfoil.
\newblock AIAA Journal \textbf{44}(3), 620--628 (2006).
\newblock \doi{10.2514/1.16658}

\bibitem{hartmann2012}
A.~Hartmann, M.~Klaas, W.~Schröder, Time-resolved stereo piv measurements of
  shock–boundary layer interaction on a supercritical airfoil.
\newblock Experiments in Fluids \textbf{52}(3), 591--604 (2012).
\newblock \doi{10.1007/s00348-011-1074-6}

\bibitem{hartmann2013a}
A.~Hartmann, M.~Klaas, W.~Schröder, Coupled airfoil heave/pitch oscillations
  at buffet flow.
\newblock AIAA Journal \textbf{51}, 1542--1552 (2013).
\newblock \doi{10.2514/1.J051512}

\bibitem{hartmann2013b}
A.~Hartmann, A.~Feldhusen, W.~Schröder, On the interaction of shock waves and
  sound waves in transonic buffet flow.
\newblock Physics of Fluids \textbf{25}(2), 026,101 (2013).
\newblock \doi{10.1063/1.4791603}

\bibitem{feldhusen2017}
A.~Feldhusen-Hoffmann, V.~Statnikov, M.~Klaas, W.~Schröder, Investigation of
  shock–acoustic-wave interaction in transonic flow.
\newblock Experiments in Fluids \textbf{59}, 15 (2017).
\newblock \doi{10.1007/s00348-017-2466-z}

\bibitem{feldhusen2021}
A.~Feldhusen-Hoffmann, C.~Lagemann, S.~Loosen, P.~Meysonnat, M.~Klaas,
  W.~Schröder, Analysis of transonic buffet using dynamic mode decomposition.
\newblock Experiments in Fluids \textbf{62}, 66 (2021).
\newblock \doi{10.1007/s00348-020-03111-5}

\bibitem{dietz2008}
G.~Dietz, H.~Mai, A.~Schröder, C.~Klein, N.~Moreaux, P.~Leconte, Unsteady
  wing-pylon-nacelle interference in transonic flow.
\newblock Journal of Aircraft \textbf{Vol. 45}, 934--944 (2008).
\newblock \doi{10.2514/6.2007-2018}

\bibitem{spinner2021}
S.~Spinner, R.~Ralf, Design of a uhbr through flow nacelle for high speed stall
  wind tunnel investigations.
\newblock Deutscher Luft- und Raumfahrtkongress  (2021).
\newblock \doi{10.25967/550043}

\bibitem{spinner2022}
S.~Spinner, R.~Rudnik, Experimental assessment of wing lower surface buffet
  effects induced by the installation of a uhbr nacelle.
\newblock CEAS Aeronautical Journal \textbf{-}, -- (2022).
\newblock \doi{10.1007/s13272-022-00632-z}

\bibitem{giannelis2017}
N.F. Giannelis, G.A. Vio, O.~Levinski, A review of recent developments in the
  understanding of transonic shock buffet.
\newblock Progress in Aerospace Sciences \textbf{92}, 39--84 (2017).
\newblock \doi{10.1016/j.paerosci.2017.05.004}

\bibitem{jacquin2009}
L.~Jacquin, P.~Molton, S.~Deck, B.~Maury, D.~Soulevant, Experimental study of
  shock oscillation over a transonic supercritical profile.
\newblock AIAA Journal \textbf{47}(9), 1985--1994 (2009).
\newblock \doi{10.2514/1.30190}

\bibitem{lintermann2020}
A.~Lintermann, M.~Meinke, W.~Schr\"oder, Zonal flow solver ({ZFS}): a highly
  efficient multi-physics simulation framework.
\newblock International Journal of Computational Fluid Dynamics
  \textbf{34}(7-8), 458--485 (2020).
\newblock \doi{10.1080/10618562.2020.1742328}

\bibitem{boris1992}
J.P. Boris, F.F. Grinstein, E.S. Oran, R.L. Kolbe, New insights into large eddy
  simulation.
\newblock Fluid Dynamics Research \textbf{10}(4-6), 199--228 (1992).
\newblock \doi{10.1016/0169-5983(92)90023-p}

\bibitem{meinke2002}
M.~Meinke, W.~Schr\"oder, E.~Krause, T.~Rister, A comparison of second- and
  sixth-order methods for large-eddy simulations.
\newblock Computers \& Fluids \textbf{31}(4), 695--718 (2002).
\newblock \doi{10.1016/S0045-7930(01)00073-1}

\bibitem{schneiders2016}
L.~Schneiders, C.~G\"unther, M.~Meinke, W.~Schr\"oder, An efficient
  conservative cut-cell method for rigid bodies interacting with viscous
  compressible flows.
\newblock Journal of Computational Physics \textbf{311}, 62--86 (2016).
\newblock \doi{10.1016/j.jcp.2016.01.026}

\bibitem{luerkens2022}
T.~L\"urkens, M.~Meinke, W.~Schr\"oder, Wall-modeled {LES} of buffet under the
  influence of engine nacelle flow.
\newblock Deutscher Luft- und Raumfahrtkongress  (2022).
\newblock \doi{10.25967/570433.}

\bibitem{fukushima2018}
Y.~Fukushima, S.~Kawai, Wall-modeled large-eddy simulation of transonic airfoil
  buffet at high reynolds number.
\newblock AIAA Journal \textbf{56}(6), 2372--2388 (2018).
\newblock \doi{10.2514/1.J056537}

\bibitem{spalding1961}
D.B. Spalding, {A Single Formula for the “Law of the Wall”}.
\newblock Journal of Applied Mechanics \textbf{28}(3), 455--458 (1961).
\newblock \doi{10.1115/1.3641728}

\bibitem{schmid2010}
P.J. Schmid, Dynamic mode decomposition of numerical and experimental data.
\newblock Journal of Fluid Mechanics \textbf{656}, 5–28 (2010).
\newblock \doi{10.1017/S0022112010001217}

\bibitem{kou2017}
J.~Kou, W.~Zhang, An improved criterion to select dominant modes from dynamic
  mode decomposition.
\newblock European Journal of Mechanics - B/Fluids \textbf{62}, 109–129
  (2017).
\newblock \doi{10.1016/j.euromechflu.2016.11.015}.
\newblock
  \urlprefix\url{https://www.sciencedirect.com/science/article/pii/S0997754616302990}

\bibitem{kou2018}
J.~Kou, S.~{Le Clainche}, W.~Zhang, A reduced-order model for compressible
  flows with buffeting condition using higher order dynamic mode decomposition
  with a mode selection criterion.
\newblock Physics of Fluids \textbf{30}(1), 016,103 (2018).
\newblock \doi{10.1063/1.4999699}

\bibitem{gao2017}
C.~Gao, W.~Zhang, J.~Kou, Y.~Liu, Z.~Ye, Active control of transonic buffet
  flow.
\newblock Journal of Fluid Mechanics \textbf{824}, 312–351 (2017).
\newblock \doi{10.1017/jfm.2017.344}

\bibitem{poplingher2019}
L.~Poplingher, D.E. Raveh, E.H. Dowell, Modal analysis of transonic shock
  buffet on 2d airfoil.
\newblock AIAA Journal \textbf{57}(7), 2851–2866 (2019).
\newblock \doi{10.2514/1.J057893}

\bibitem{jovanovic2014}
M.R. Jovanović, P.J. Schmid, J.W. Nichols, Sparsity-promoting dynamic mode
  decomposition.
\newblock Physics of Fluids \textbf{26}(2), 024,103 (2014).
\newblock \doi{10.1063/1.4863670}.
\newblock \urlprefix\url{https://doi.org/10.1063/1.4863670}.
\newblock
  {\href{https://arxiv.org/abs/https://doi.org/10.1063/1.4863670}{{https://doi.org/10.1063/1.4863670}}}

\bibitem{schauerte2022}
C.J. Schauerte, A.M. Schreyer, Experimental analysis of transonic buffet
  conditions on a two-dimensional supercritical airfoil (under review).
\newblock AIAA Journal  (2022)

\bibitem{zauner2020}
M.~Zauner, N.D. Sandham, Wide domain simulations of flow over an unswept
  laminar wing section undergoing transonic buffet.
\newblock Physical Review Fluids \textbf{5}, 083,903 (2020).
\newblock \doi{10.1103/PhysRevFluids.5.083903}

\bibitem{moise2021}
P.~Moise, M.~Zauner, N.D. Sandham, Large eddy simulations and modal
  reconstruction of laminar transonic buffet (under revision).
\newblock Journal of Fluid Mechanics  (2021)

\bibitem{kawai2012}
S.~Kawai, J.~Larsson, Wall-modeling in large eddy simulation: Length scales,
  grid resolution, and accuracy.
\newblock Physics of Fluids \textbf{24}(1), 015,105 (2012).
\newblock \doi{10.1063/1.3678331}

\bibitem{schlatter2012}
P.~Schlatter, R.~Örlü, Turbulent boundary layers at moderate reynolds
  numbers: inflow length and tripping effects.
\newblock Journal of Fluid Mechanics \textbf{710}, 5–34 (2012).
\newblock \doi{10.1017/jfm.2012.324}

\bibitem{ewert2003}
R.~Ewert, W.~Schröder, Acoustic perturbation equations based on flow
  decomposition via source filtering.
\newblock Journal of Computational Physics \textbf{188}(2), 365–398 (2003).
\newblock \doi{10.1016/S0021-9991(03)00168-2}

\bibitem{moise2022}
P.~Moise, M.~Zauner, N.~Sandham, S.~Timme, W.~He, Transonic buffet
  characteristics under conditions of free and forced transition.
\newblock AIAA Journal pp. 1--16 (2022).
\newblock \doi{10.2514/1.J062362}

\end{thebibliography}
%% if required, the content of .bbl file can be included here once bbl is generated
%%\input sn-article.bbl

%% Default %%
%%\input sn-sample-bib.tex%

\end{document}